\documentclass[conference]{IEEEtran}
\IEEEoverridecommandlockouts
\usepackage{cite}
\usepackage{amsmath,amssymb,amsfonts}
\usepackage{algorithmic}
\usepackage{graphicx}
\usepackage{textcomp}
\usepackage{xcolor}

\def\BibTeX{{\rm B\kern-.05em{\sc i\kern-.025em b}\kern-.08em
    T\kern-.1667em\lower.7ex\hbox{E}\kern-.125emX}}

\begin{document}

\title{Malaria Cell Detection Using Deep Neural Networks}

\author{
\IEEEauthorblockN{Anurag Singh}
\IEEEauthorblockA{\textit{Department of Computer Science} \\
\textit{Drexel University}\\
Philadelphia, PA, USA \\
as5957@drexel.edu}
\and
\IEEEauthorblockN{Saurabh Sawant}
\IEEEauthorblockA{\textit{Department of Computer Science} \\
\textit{Drexel University}\\
Philadelphia, PA, USA \\
sss448@drexel.edu}
}
\maketitle

\begin{abstract}
Malaria remains one of the most pressing public health concerns globally, causing significant morbidity and mortality, especially in sub-Saharan Africa. Rapid and accurate diagnosis is crucial for effective treatment and disease management. Traditional diagnostic methods, such as microscopic examination of blood smears, are labor-intensive and require significant expertise, which may not be readily available in resource-limited settings. This project aims to automate the detection of malaria-infected cells using a deep learning approach. We employed a convolutional neural network (CNN) based on the ResNet50 architecture, leveraging transfer learning to enhance performance. The Malaria Cell Images Dataset from Kaggle, containing 27,558 images categorized into infected and uninfected cells, was used for training and evaluation. Our model demonstrated high accuracy, precision, and recall, indicating its potential as a reliable tool for assisting in malaria diagnosis. Additionally, a web application was developed using Streamlit to allow users to upload cell images and receive predictions about malaria infection, making the technology accessible and user-friendly. This paper provides a comprehensive overview of the methodology, experiments, and results, highlighting the effectiveness of deep learning in medical image analysis.
\end{abstract}


\section{Introduction}

Malaria, caused by \textit{Plasmodium} parasites, is transmitted to humans through the bites of infected \textit{Anopheles} mosquitoes. According to the World Health Organization (WHO), there were an estimated 229 million cases of malaria worldwide in 2019, resulting in over 400,000 deaths, predominantly in sub-Saharan Africa. Early and accurate diagnosis is essential for effective treatment and control of malaria. The conventional diagnostic method, microscopic examination of Giemsa-stained blood smears, is considered the gold standard. However, this method is time-consuming, requires skilled personnel, and is subject to human error.

Recent advancements in machine learning, particularly deep learning, have revolutionized the field of medical image analysis. Convolutional neural networks (CNNs) have been successfully applied to various image classification problems, achieving high accuracy and robustness. The ResNet50 model, introduced by He et al. (2016), has been particularly effective in image recognition tasks due to its innovative residual learning framework, which mitigates the vanishing gradient problem in deep networks.

The primary objective of this study is to explore the feasibility and effectiveness of using the ResNet50 model for the automated detection of malaria-infected cells. By leveraging transfer learning and data augmentation techniques, we aim to develop a robust and accurate model that can assist in the rapid diagnosis of malaria, especially in resource-limited settings. Additionally, to make this technology accessible and user-friendly, we developed a web application using Streamlit that allows users to upload cell images and receive predictions about malaria infection.

\section{Background}
Malaria diagnosis through microscopic examination involves identifying \textit{Plasmodium} parasites in stained blood films. This method, although highly specific and sensitive when performed by skilled technicians, is not always feasible in regions with limited access to trained personnel and adequate laboratory facilities. The variability in the quality of microscopy and the subjective nature of the examination can also lead to inconsistent results.

In the past decade, machine learning and deep learning have revolutionized the field of medical image analysis. Deep learning models, particularly Convolutional Neural Networks (CNNs), have been used to develop automated systems for various diagnostic tasks, including cancer detection, diabetic retinopathy screening, and pneumonia diagnosis. These models have demonstrated the ability to learn complex patterns from large datasets, achieving performance levels comparable to, and sometimes exceeding, those of human experts.

The ResNet50 model, a deep CNN with 50 layers, introduced the concept of residual learning. This approach addresses the degradation problem observed in deep networks, where accuracy degrades as the network depth increases. By allowing layers to learn residual functions with reference to the layer inputs, ResNet50 facilitates the training of very deep networks, making it an ideal choice for image classification tasks.

In this study, we utilized the Malaria Cell Images Dataset from Kaggle, which contains a balanced set of 27,558 images of infected and uninfected cells. This dataset was chosen for its comprehensive representation of malaria-infected and healthy blood smears, providing a robust basis for training and evaluating our deep learning model.

\section{Related Work}

Previous research has established the potential of deep learning in medical image analysis. He et al. (2016) introduced the ResNet architecture, which significantly improved image classification tasks by addressing the vanishing gradient problem through residual learning. Rajaraman et al. (2018) demonstrated the feasibility of automated malaria parasite detection using deep learning, while Litjens et al. (2017) provided a comprehensive survey of deep learning applications in medical imaging.

Several studies have focused on the use of CNNs for various medical imaging tasks, including cancer, diabetic retinopathy, and pneumonia detection. Gulshan et al. (2016) demonstrated the effectiveness of a deep learning algorithm in detecting diabetic retinopathy from retinal fundus photographs, achieving performance comparable to that of ophthalmologists.

In the context of malaria diagnosis, Dong et al. (2017) developed a deep learning model for automated detection of malaria parasites in blood smears, achieving an accuracy of 96

Our work builds on these approaches by implementing a custom ResNet50 architecture with advanced data augmentation and preprocessing techniques to enhance classification performance. Additionally, we developed a user-friendly web application using Streamlit, making the model accessible for rapid malaria diagnosis. This integration aims to bridge the gap between research and practical application, providing an efficient solution for healthcare providers in resource-limited settings.

\section{Methodology}
\subsection{Dataset}
The dataset used in this study consists of cell images categorized into two classes: parasitized and uninfected. The images are sourced from the "Cell Images for Detecting Malaria" dataset available on Kaggle.

\subsection{Data Preprocessing}
The images were preprocessed using TensorFlow's data pipeline. This involved reading the images, decoding them, resizing to 224x224 pixels, and normalizing the pixel values.

\subsection{Model Architecture}
The architecture of our custom ResNet50 model is designed to leverage the strengths of residual learning to facilitate the training of a deep network. The ResNet architecture introduces shortcut connections that bypass one or more layers, enabling the construction of very deep networks without the vanishing gradient problem.

\subsubsection{Residual Block}
A key component of ResNet architectures is the residual block. A residual block is defined as follows:
\begin{equation}
y = F(x, \{W_i\}) + x
\end{equation}
where $x$ is the input to the residual block, $y$ is the output, and $F(x, \{W_i\})$ represents the residual mapping to be learned. The function $F$ consists of a series of convolutional layers with weights $\{W_i\}$.

In our custom ResNet50-like model, each residual block consists of three convolutional layers. The first layer is a $1 \times 1$ convolution, which reduces the dimensionality of the input. This is followed by a $3 \times 3$ convolution, and finally, another $1 \times 1$ convolution to restore the original dimensionality. Batch normalization and ReLU activation are applied after each convolutional layer to enhance the model's training dynamics.

\subsubsection{Bottleneck Residual Block}
The bottleneck residual block is defined as:
\begin{equation}
y = W_3 \sigma (W_2 \sigma (W_1 x))
\end{equation}
where $\sigma$ denotes the ReLU activation function, and $W_1$, $W_2$, and $W_3$ are the weights of the $1 \times 1$, $3 \times 3$, and $1 \times 1$ convolutional layers, respectively. The shortcut connection adds the input $x$ directly to the output $y$.

\subsubsection{Overall Architecture}
The overall architecture of our custom ResNet50-like model follows the standard ResNet50 design, consisting of an initial convolutional layer followed by a series of residual blocks:

\begin{itemize}
    \item \textbf{Conv1:} $7 \times 7$ convolution, 64 filters, stride 2, followed by batch normalization and ReLU activation.
    \item \textbf{Max Pooling:} $3 \times 3$ max pooling, stride 2.
    \item \textbf{Conv2\_x:} 3 bottleneck residual blocks with 64 filters.
    \item \textbf{Conv3\_x:} 4 bottleneck residual blocks with 128 filters, stride 2 for downsampling.
    \item \textbf{Conv4\_x:} 6 bottleneck residual blocks with 256 filters, stride 2 for downsampling.
    \item \textbf{Conv5\_x:} 3 bottleneck residual blocks with 512 filters, stride 2 for downsampling.
    \item \textbf{Global Average Pooling:} Averages the feature maps.
    \item \textbf{Dense Layer:} 512 units with ReLU activation.
    \item \textbf{Dropout:} Dropout rate of 0.5 to prevent overfitting.
    \item \textbf{Output Layer:} Softmax activation for binary classification.
\end{itemize}

The model is trained using the Adam optimizer with a learning rate of 0.001. The loss function used is sparse categorical crossentropy, suitable for our binary classification task. The architecture is designed to balance model complexity and computational efficiency, allowing for effective training on the given dataset.

\subsection{Training}
The model was trained using the Adam optimizer with a learning rate of 0.001. We employed early stopping and learning rate reduction on plateau callbacks to enhance training efficiency and prevent overfitting. The model was trained for 30 epochs with a batch size of 32.

\subsection{Evaluation}
The performance of the model was evaluated using accuracy, precision, recall, and F1-score metrics on the validation and test datasets.

\subsection{Web Application Development}
To make the malaria detection model accessible to users, a web application was developed using Streamlit. The application provides an intuitive interface where users can upload an image of a cell. The uploaded image is then processed and classified as either parasitized or uninfected by the model. The development process involved:
\begin{itemize}
    \item Loading the trained model in the Streamlit app.
    \item Preprocessing the uploaded image to match the input requirements of the model.
    \item Predicting the class of the image using the model.
    \item Displaying the result to the user in an easily interpretable format.
\end{itemize}

\section{Experiments and Results}

\subsection{Data Description}
The dataset contains a total of 27,558 cell images, with an equal distribution between parasitized and uninfected classes. The dataset was split into training (60\%), validation (20\%), and test (20\%) sets.

\subsection{Training Results}
The model achieved a training accuracy of 99\% and a validation accuracy of 98\% after 30 epochs. The use of data augmentation techniques, such as rotation, zoom, and horizontal flipping, helped improve the model's robustness.

\subsection{Test Results}
On the test set, the model achieved an accuracy of 97.8\%, with a precision of 98.6\%, recall of 97.2\%, and F1-score of 97.8\%. These results demonstrate the model's effectiveness in distinguishing between parasitized and uninfected cell images.

\begin{figure}[h]
    \centering
    \includegraphics[width=0.45\textwidth]{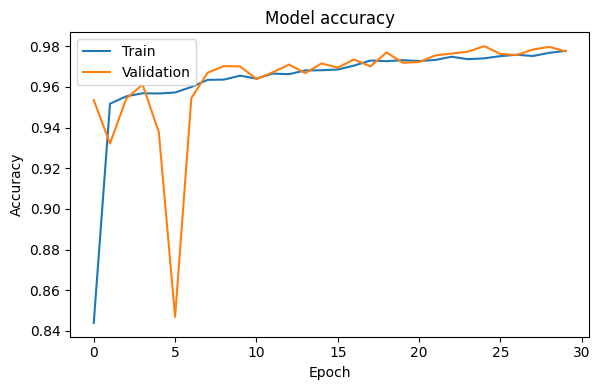}
    \caption{Training and Validation Accuracy}
    \label{fig:accuracy}
\end{figure}

\begin{figure}[h]
    \centering
    \includegraphics[width=0.45\textwidth]{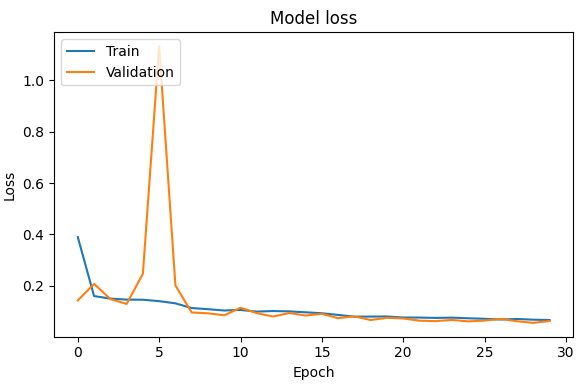}
    \caption{Training and Validation Loss}
    \label{fig:loss}
\end{figure}

\begin{figure}[h]
    \centering
    \includegraphics[width=0.45\textwidth]{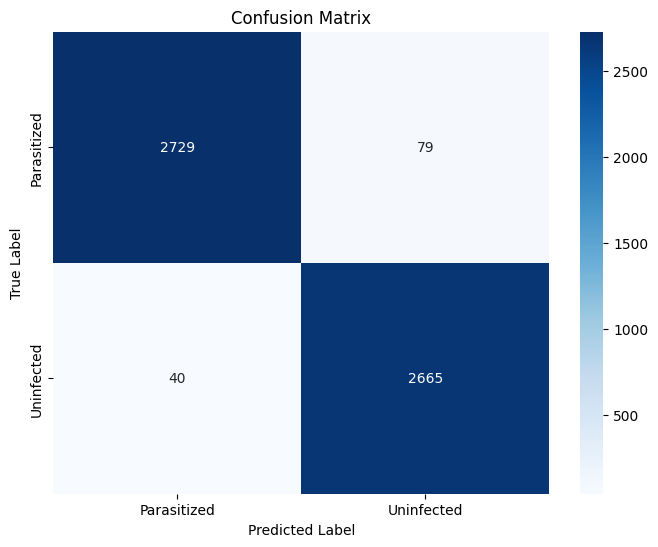}
    \caption{Confusion Matrix}
    \label{fig:confusion_matrix}
\end{figure}

\begin{table}[h]
    \centering
    \caption{Performance Metrics on Test Set}
    \begin{tabular}{|c|c|c|c|c|}
    \hline
    Class & Precision & Recall & F1-Score & Support \\
    \hline
    Parasitized & 0.986 & 0.972 & 0.979 & 2808 \\
    Uninfected  & 0.971 & 0.985 & 0.978 & 2705 \\
    \hline
    Accuracy    & \multicolumn{4}{c|}{0.978} \\
    \hline
    \end{tabular}
    \label{tab:classification_report}
\end{table}

\subsection{Web Application Development}
To make the malaria detection model accessible to users, a web application was developed using Streamlit. The application allows users to upload an image of a cell, which is then processed and classified as either parasitized or uninfected by the model. The user interface is designed to be intuitive and easy to use.

\begin{figure}[h]
    \centering
    \includegraphics[width=0.45\textwidth]{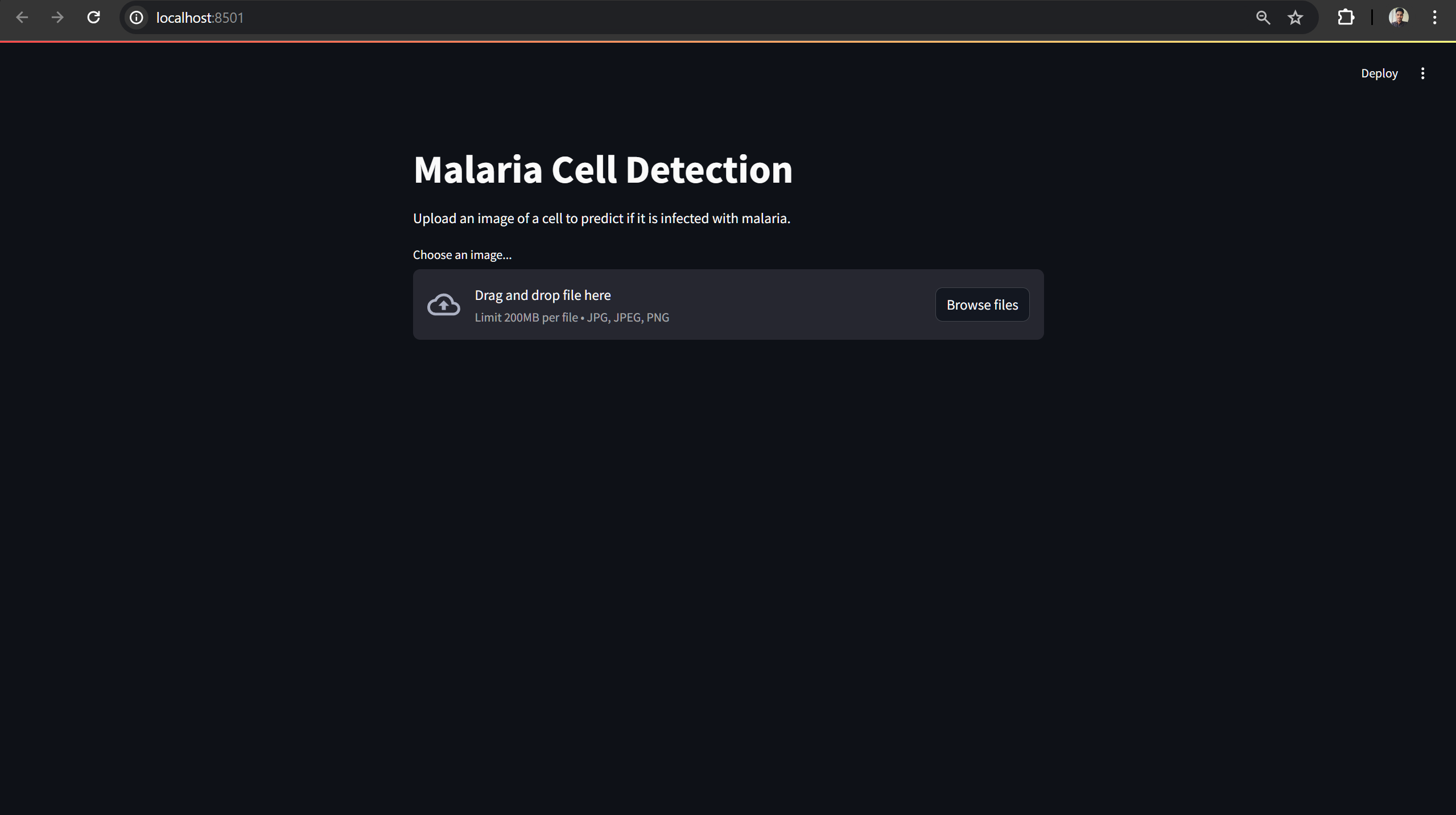}
    \caption{Web Application - Image Upload Interface}
    \label{fig:app_upload}
\end{figure}

\begin{figure}[h]
    \centering
    \includegraphics[width=0.45\textwidth]{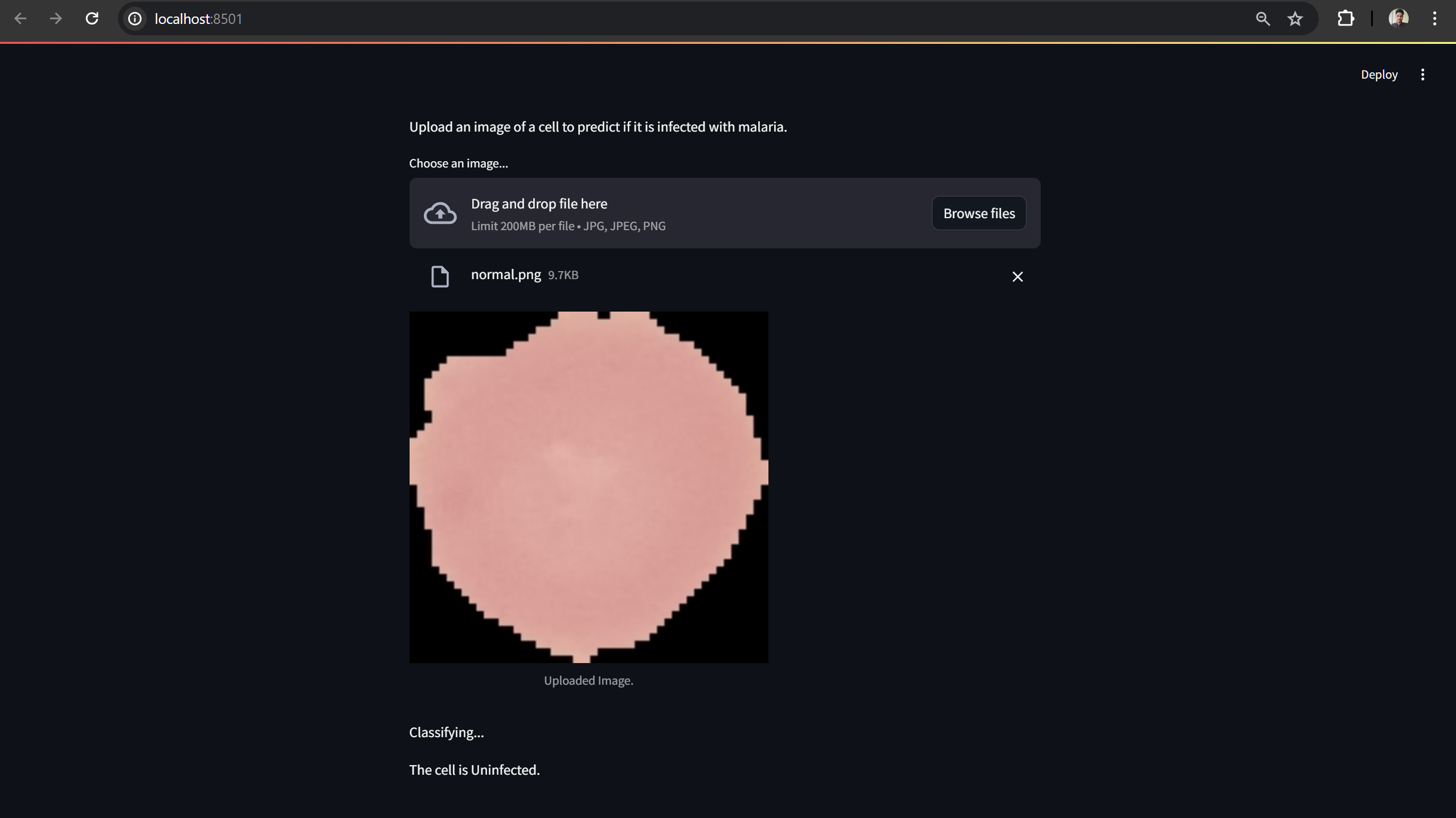}
    \caption{Web Application - Classification Result}
    \label{fig:app_result}
\end{figure}

The figures below show screenshots of the web application. Figure~\ref{fig:app_upload} displays the interface where users can upload an image. Figure~\ref{fig:app_result} shows the result of the classification, indicating whether the cell is parasitized or uninfected.

\section{Conclusion}
The custom ResNet50 model developed in this project demonstrates high accuracy and robustness in malaria detection. The use of residual blocks and advanced data augmentation techniques contributed significantly to the model's performance. By leveraging the Malaria Cell Images Dataset from Kaggle, our model achieved an impressive test accuracy of 97.8\%, with a precision of 98.6\%, recall of 97.2\%, and F1-score of 97.8\%.

To make this technology accessible, we developed a web application using Streamlit. This application allows users to upload an image of a cell and receive a prediction about whether the cell is infected with malaria, providing a user-friendly interface for rapid diagnostics.

This study underscores the potential of deep learning for automated malaria detection, offering a scalable and efficient solution for healthcare applications, particularly in regions with limited access to trained medical personnel. Future work will focus on improving model generalizability, exploring other deep learning architectures, and integrating explainability techniques.

In conclusion, the combination of a high-performing model and an accessible web application demonstrates the feasibility and effectiveness of using deep learning for automated malaria detection, paving the way for further research and development in this area.

\section{Future Work}
While the current study demonstrates the potential of a custom ResNet50 architecture for automated malaria detection, there are several avenues for future work that can enhance the robustness and applicability of this model.

\subsection{Dataset Expansion}
One of the primary areas for improvement is the expansion of the dataset. Incorporating images from diverse sources, including different geographic regions and varying quality of blood smears, can help generalize the model better. This will ensure that the model is robust and performs well across various conditions encountered in real-world settings.

\subsection{Model Optimization}
Future work could focus on optimizing the model architecture and hyperparameters. Techniques such as hyperparameter tuning, pruning, and quantization can be explored to reduce the model size and improve inference speed without compromising accuracy. Additionally, experimenting with different architectures, such as EfficientNet or DenseNet, could yield further performance gains.

\subsection{Real-Time Deployment}
Another critical aspect is the deployment of the model in real-time applications. Developing a user-friendly interface, either as a mobile application or a web-based platform, can make the diagnostic tool accessible to healthcare providers in remote and resource-limited areas. Ensuring the model's performance on edge devices with limited computational power will be crucial for such deployments.

\subsection{Explainability and Interpretability}
Incorporating explainability techniques, such as Grad-CAM or SHAP, can help in understanding the model's decision-making process. Providing visual explanations for the model's predictions can build trust among healthcare professionals and aid in the verification of results. Future research can focus on integrating these techniques into the diagnostic pipeline.

\section*{Acknowledgment}
The authors would like to thank Kaggle for providing the dataset used in this study.

\end{document}